\documentstyle[aps,eqsecnum,preprint]{revtex}
\tightenlines
\begin{document}

\title{A hidden symmetry of collisionless sound \\
       of Bose condensates in anisotropic traps}
\author{Martin Fliesser and Robert Graham}
\address{Fachbereich Physik, Universit\"at-Gesamthochschule Essen\\
45117 Essen,\, Germany}
\maketitle

\begin{abstract}
We derive a symmetry property 
for the Fourier-transform
of the collisionless sound modes
of Bose condensates in anisotropic traps
connected with a somewhat hidden conservation law.
We discuss its possible observation
by dispersive light scattering.
\end{abstract}

\vskip1pc


\vskip2pc
\narrowtext

Since the achievement of Bose-Einstein condensates 
in trapped alkali gases
\cite{anderson,bradley,davis} at ultra-low temperatures
numerous aspects of these systems have been investigated
both from the experimental and the theoretical side.
Adding a time dependent perturbation 
to the trap potential
the lowest frequencies of the density oscillation spectrum
were measured 
by observing condensate shape oscillations 
either by time of flight measurements \cite{4,5,6}
or by dispersive light scattering \cite{7}.
In this paper we want to 
investigate these mode functions in detail
and show how they reflect a special symmetry of the system.
We disuss how this symmetry property could
be observed in phase-contrast imaging measurements \cite{7,8}
and comment on the feasibility of its observation
in inelastic light sacttering.

We restrict ourselves 
to the density fluctuations 
of energies much smaller than the chemical potential
($\hbar\omega\ll\mu$)
of a trapped Bose condensate
at temperature $T=0$.
The equations governing these collisionless sound modes
in a weakly interacting system in mean field approximation
can be written in the form
of hydrodynamic equations \cite{stringari,griffin}.
For isotropic \cite{stringari}, 
axially symmetric \cite{fliesser}
and even fully anisotropic harmonic traps \cite{csordas} 
the solutions of these equations can be classified 
in terms of three quantum numbers.
We consider these mode functions in Fourier space
where we scale all momenta 
by $k_i=\tilde k_i\,\sqrt{m\omega_i^2/2\mu}$
for $i=x,y,z$
with chemical potential $\mu$
and trap frequency $\omega_i$ in direction $i$.

Before going into any technical details we state the central result
of this paper and discuss how it may be tested experimentally:
We show in this paper that the Fourier transform $\tilde\phi(\bf{\tilde k})$
of the mode functions for traps of {\it arbitrary anisotropy}
factorizes into a radial part 
depending only on the {\it modulus} of the scaled wave vector
\begin{eqnarray}
   \tilde k = |{\bf\tilde k}| = \sqrt{2\mu/m
   (k_x^2/\omega_x^2 + k_y^2/\omega_y^2 + k_z^2/\omega_z^2)}
     \label{modulus}
\end{eqnarray}
times an angular part $g(\theta_k,\phi_k)$ 
depending only on the two angles $\theta_k, \phi_k$ of
spherical coordinates
in scaled Fourier space.
Furthermore the radial part 
does not depend on the trap anisotropy anymore and
can be determined explicitly.
We simply have
\begin{eqnarray}
    \tilde\phi({\bf \tilde k})
   = \frac{j_{\nu}(\tilde k)}{\tilde k}
     \,\times\,
     g(\theta_k,\phi_k)
     \label{result}\,,
\end{eqnarray}
where $j_{\nu}$ denotes the spherical Bessel-function,
the positive integer degree $\nu$ depending on the three quantum numbers
of the mode functions, which will be specified below. 
This factorization is remarkable since neither the original problem
nor the problem with rescaled momenta are spherically symmetric.
 In fact it follows
from a somewhat hidden conservation law of an operator with spherical
symmetry after rescaling
as we shall show below.

Before we derive (\ref{result})
we wish to illustrate this property
by discussing how it 
could be observed in an experiment.
One possibility to observe density fluctuations is
by the direct observation of a condensate
by phase-contrast imaging \cite{7,8}.
A beam of light
off resonace with any atomic transition
accumulates a phase shift on its way through the condensate,
which is proportional to the density of atoms 
integrated along its optical path.
In phase-contrast imaging the direct beam of light is shifted by
$\pi/2$
by passing it through a $\lambda/4$-plate 
in the focal plane.
In the interference pattern
between the direct beam and the deflected light on a screen
phase differences are thereby turned into amplitude differences.
Since typical angles of deflection 
in the  experiment of \cite{8} are only of the order of several mrad
we can approximate the paths of deflected light also as straight lines
parallel to the optical axis.
The observed amplitude differences are therefore proportional
to the projection of the three dimensional number density 
of the condensate
on a two dimensional screen.
By this technique it is possible to take time resolved sequences
of pictures of an oscillating condensate \cite{7},
after its excitation via a modulation of the trap 
with a certain frequency.

By Fourier transformimg these sequences of pictures in time 
with the known mode-frequency
one gets the projection of the density fluctuation mode 
$\phi({\bf r})$
onto the screen, i.e.
the mode function integrated along 
the direction of the optical axis.
The Fourier transform
 $\tilde\phi({\bf k})$ of the mode function
for any wave vector $\bf k$ perpendicular to the optical axis
can then be obtained by taking the spatial Fourier transformation of
this picture with the wave vector $\bf k$ in the image-plane.
To test relation (\ref{result}) the Fourier transform must be taken
for a range of $k$-values with fixed direction of $\bf k$,
and thus also of the scaled $\bf\tilde k$,
calculating the modulus $\tilde k$ from (\ref{modulus}).
The chemical potential neccessary for the application of
(\ref{modulus}) must be determined from the number of 
atoms in the trap.
For fixed direction of ${\bf \tilde k}$
$g(\theta_k,\phi_k)$ only occurs as a constant factor,
which of course depends on the direction chosen
and which may vanish in some directions.

The scaling introduced for the momenta
corresponds to scaling all coordinates
with the Thomas-Fermi radius in this direction
according to $r_i=\tilde r_i\,\sqrt{2\mu/m\omega_i^2}$
for $i=x,y,z$.
By this spatial rescaling an anisotropic harmonic trap 
is transformed to an isotropic one
and pictures of elliptical condensates
are transformed to spherical ones.
Therefore, instead of calculating the spatial Fourier transform 
of pictures of mode functions in scaled momenta
one can also first unfold these pictures
to spherical symmetry
and then apply normal Fourier transformation.
In an experiment the optical axis probably would be oriented 
along one axis of the trap, e.g. the $y$-axis.
The simplest case then certainly is to take 
the orientation of $\bf\tilde k$ either along the z-axis,
with $\tilde k = k_z \sqrt{2\mu/m\omega_z^2}$,
or along the x-axis
with $\tilde k = k_x \sqrt{2\mu/m\omega_x^2}$.
For an arbitrary, but fixed orientation of $\bf\tilde k$
we have to keep the ratio $k_x/k_z$ constant 
with the modulus given by (\ref{modulus}).
In summary, once having taken a sequence of phase-contrast images
for a particular mode,
relation (\ref{result}) could be tested easily 
by a simple analysis.

Now let us derive the property (\ref{result}).
For a Bose gas 
below the Bose-Einstein transition
the usual separation of the particle field operator
into a classical part $\phi_0({\bf r})$,
the condensate wave function, and a residual operator
$\hat\phi({\bf r})$, describing excitations out of the condensate,
yields for the number-density operator
\begin{eqnarray}
   \hat\rho({\bf r}) 
   =  |\phi_0({\bf r})|^2 
   + \Big(\phi_0^*({\bf r})\hat\phi({\bf r}) 
   +  \phi_0({\bf r})\hat\phi^{\dagger}({\bf r})\Big)
   +  \hat\phi^{\dagger}({\bf r}) \hat\phi({\bf r})  
      \nonumber \,.
\end{eqnarray}
Thermal averaging of the last term gives 
the condensate depletion at $T=0$ 
plus the density of thermally excited atoms.
At low temperatures for a weakly interacting system
the condensate density $|\phi_0({\bf r})|^2$ 
contains most of the particles,
and
the second term 
dominates by far the last term.
Restricting ourselves 
to the low-lying density fluctuations 
with $\hbar\omega\ll\mu$
\cite{stringari},
one single mode of these dominant density fluctuations 
with mode frequency $\omega$ takes the form
\cite{griffin,fetter1}
\begin{eqnarray}
      \phi_0^*({\bf r})\hat\phi({\bf r}) 
   +  \phi_0({\bf r})\hat\phi^{\dagger}({\bf r})
   = i\sqrt{\frac{\hbar\omega}{2g}}
    (\phi({\bf r})\hat\alpha e^{-i\omega t} 
   - \phi^*({\bf r})\hat\alpha^{\dagger} e^{i\omega t})
   \nonumber\,.
\end{eqnarray}
Here $g$ is the atom-atom interaction constant $4\pi\hbar^2 a/m$,
with the s-wave scattering length $a$,
and 
$\phi({\bf r})$
a normalized density fluctuation mode. 
These density fluctuations 
and the corresponding velocity fluctuations
can be described by hydrodynamic equations
of the Euler type. From this description one
can derive 
a wave equation
\cite{stringari,griffin}
\begin{eqnarray}
  -\omega^2\,\phi({\bf r}) 
  ={\bf\nabla}\cdot\frac{g}{m}|\phi_0({\bf r})|^2
   {\bf\nabla}\phi({\bf r})
   \label{wave.equation}
\end{eqnarray}
which both the normalized density modes
and the velocity fluctuation modes
have to fulfill
\cite{stringari,griffin}.

Features of Bose condensates are most readily observable
in the limit of large condensates.
In this limit
the Thomas-Fermi approximation for the condensate wave function
$g|\phi_0({\bf r})|^2=(\mu-U({\bf r}))\Theta(\mu-U({\bf r}))$
is justified \cite{baym},
which restricts solutions of (\ref{wave.equation})
to the interior of the Thomas-Fermi surface $\mu=U({\bf r})$.
For condensates in isotropic harmonic traps
the spectrum and the eigenfunctions 
of (\ref{wave.equation})
were calculated by Stringari \cite{stringari}.
They depend on a radial and an angular quantum number $n_r$ and $l$,
the spectrum being degenerate 
with respect to the azimuthal quantum number $m_l$.

The atom-traps used in the different realisations 
of Bose condensates in alkali atoms
can be decribed in very good approximation
by an anisotropic harmonic potential
$U({\bf r})=m(\omega_x^2 x^2 +\omega_y^2 y^2 +\omega_z^2 z^2)/2$.
Here we do not restrict ourselves to any axial or spherical symmetry.
Due to the anisotropy of the potential new features show up:
the azimuthal quantum number $m_l$ is no longer a good quantum
number,
at energies of the order of $\mu$
the system ceases to be strictly integrable, and
a large chaotic component can be seen in the classical phase space 
\cite{fliesser.class}.
However in the limit of low energies ($\hbar\omega\ll\mu$) 
the systems reduces to an integrable one \cite{csordas}
and an additional constant of motion can be identified
\cite{fliesser}
\begin{eqnarray}
   \hat B =
   - \sum_i \frac{2\mu}{m\omega_i^2}(\frac{\partial}{\partial r_i})^2
   + ({\bf r}\cdot{\bf\nabla})^2 + 3 {\bf r}\cdot{\bf\nabla}
     \label{const}\,.
\end{eqnarray}
We shall now show
how (\ref{result})
follows from this conserved quantity.

Let us introduce a scaled position-vector
${\bf\tilde r}$ with dimensionless components $\tilde r_i$
all in the range $[0,1]$
by the scaling $r_i=\tilde r_i\,\sqrt{2\mu/m\omega_i^2}$
for $i=x,y,z\,$.
Any harmonic trap potential now becomes
$U({\bf\tilde r})=\mu\tilde r^2$
and the Thomas-Fermi surface
is mapped to the unit sphere.
The wave equation (\ref{wave.equation}) 
takes the form
\begin{eqnarray}
  - \omega^2\,\phi({\bf\tilde r}) 
  = \frac{1}{2}\sum_i \Big[
    \omega_i^2 \frac{\partial}{\partial\tilde r_i}
    (1-\tilde r^2)
   \frac{\partial}{\partial\tilde r_i}
   \Big]\,\phi({\bf\tilde r})
   \nonumber\,.
\end{eqnarray}
Now let us change to scaled Fourier space.
Since terms from partial integration
don't occur due to the vanishing of the condensate wave function
at the condensate boundary
we simply can replace
$\tilde r_j \rightarrow i\partial/\partial\tilde k_j$
and
$\partial/\partial\tilde r_j \rightarrow ik_j$.
The modulus of the scaled momentum $\tilde k$ 
is given by (\ref{modulus}).
After simple rearrangements we get from (\ref{wave.equation})
\begin{eqnarray}
  - \omega^2\,\tilde\phi({\bf\tilde k}) 
  = \sum_i \Big[ \frac{1}{2} 
    \omega_i^2 \tilde k_i^2
    (1+{\bf \nabla_{\tilde k}}^2)
  + \omega_i^2 \tilde k_i \frac{\partial}{\partial\tilde k_i}
    \Big]\,\tilde\phi({\bf\tilde k})
    \nonumber\,.
\end{eqnarray}
Next we introduce spherical coordinates
in scaled Fourier space
in the usual way as 
$\tilde k_x = \tilde k \sin\theta_k\cos\phi_k\,,
 \tilde k_y = \tilde k \sin\theta_k\sin\phi_k$
and $\tilde k_z = \tilde k \cos\theta_k \,$.
The  explicit form of the wave equation then becomes
\begin{eqnarray}
  - \omega^2\,\tilde\phi(\tilde k,\theta_k,\phi_k) 
  &=&  \frac{1}{2}\bigg( 
    \Big(\omega^2_x \sin^2\theta_k\cos^2\phi_k 
     +\omega^2_y \sin^2\theta_k\sin^2\phi_k 
     + \omega^2_z \cos^2\phi_k\Big)
     \nonumber \\
   &\times& \bigg[ \tilde k^2 
         + \tilde k^2 (\frac{\partial}{\partial\tilde k})^2
         + 4 \tilde k \frac{\partial}{\partial\tilde k}
         + (\frac{\partial}{\partial\theta_k})^2
         + \cot\theta_k \frac{\partial}{\partial\theta_k}
         + \frac{1}{\sin^2\theta_k}(\frac{\partial}{\partial\phi_k})^2
\bigg]
    \nonumber \\
    &+&\sin2\theta_k \Big(\omega^2_x \cos^2\phi_k 
    +\omega^2_y \sin^2\phi_k - \omega^2_z\Big)
      \frac{\partial}{\partial\theta_k}
     -\sin2\phi_k\Big(\omega^2_x - \omega^2_y\Big)
      \frac{\partial}{\partial\phi_k} \bigg)
      \,\tilde\phi(\tilde k,\theta_k,\phi_k)
    \nonumber\,.
\end{eqnarray}
We see that the radial part, 
depending on $\tilde k$ only,
can be separated from this equation
which justifies the factorized form of the
separation ansatz (\ref{result}).
Our point now is 
that the  separated equation for the radial part
simply is the eigenvalue equation
for the operator $\hat B$
in the scaled Fourier representation.
To see this let us transform $\hat B$
 to the ${\bf \tilde k}$-representation.
Starting in coordiante space again,
scaling the coordinates in (\ref{const}) first,
and then changing to Fourier representation
we get
\begin{eqnarray}
   \hat B
  &=& -{\bf\tilde\nabla}^2 
   + ({\bf\tilde r}\cdot{\bf\tilde\nabla})^2 
   + 3 {\bf\tilde r}\cdot{\bf\tilde\nabla}
    \nonumber \\
  &=& \tilde k^2 
   +  \tilde k^2 (\frac{\partial}{\partial\tilde k})^2
   +  4 \tilde k \frac{\partial}{\partial\tilde k}
     \label{const.scal}\,,
\end{eqnarray}
which is indeed precisely the operator
occuring in the radial part of the wave equation.
Remarkably the eigenvalue-equation of $\hat B$
does not depend on
the frequencies $\omega_i$ anymore,
its spectrum and its solutions are the same for arbitrary
anisotropies.

The eigenfunctions of $\hat B$ in the ${\bf \tilde k}$-representation
can be found in terms of spherical Bessel-functions $j_{\nu}$
\begin{eqnarray}
    \hat B \,\, \frac{j_{\nu}(\tilde k)}{\tilde k}
   = (\nu-1)(\nu+2)\,\frac{j_{\nu}(\tilde k)}{\tilde k}
    \nonumber\,.
\end{eqnarray}
The spectrum of $\hat B$ was previously derived in \cite{fliesser},
however, without stating the eigenfunctions.
In the notation of \cite{fliesser} for axially symmetric traps
one has $\nu-1=n+m_l\,$, where $m_l$ denotes the azimuthal quantum
number.
Previous solutions
of (\ref{wave.equation})
started with a polynomial ansatz in Cartesian coordinates
\cite{stringari,oehberg,csordas} for $\phi({\bf r})\,$.
We note that $\nu-1$
is simply the highest degree of such a polynomial solution.
For the general anisotropic case the 
dependence of $\nu$ on the three quantum numbers is given in \cite{csordas}.
In terms of the quantum numbers of the isotropic problem 
\cite{stringari}
one has $\nu-1=2n_r+l\,$.
For this isotropic case (\ref{result})
was already derived by Fetter \cite{fetter2}
by direct Fourier transformation of the spatial mode functions.
Since the conserved quantity $\hat B$
does not depend on the trap frequencies $\omega_x, \omega_y, \omega_z$,
subspaces of the Hilbert space for different values 
of the quantum number $\nu$
remain orthogonal for all anisotropies.
So the same dependence on the modulus of the momentum $\tilde k$
as for the isotropic case
holds for all anisotropies.


Let us remark now on the range of validity of our discussion
which is limited by our use of
the approximate low-frequency form (\ref{wave.equation})
of the wave equation and the Thomas-Fermi approximation.
For the description by the wave equation (\ref{wave.equation})
we have to require $\hbar\omega \ll \mu$.
A second condition arises from the finite
width of the boundary layer
$l_i=(2\mu/m\omega_i^2)^{1/2}(\hbar\omega_i/4\mu)^{2/3}$
for the directions $i=x,y,z$,
limiting the momenta ${\bf k}$ for which the 
Thomas-Fermi approximation can still be used: There is
a maximum momentum $k_{li}=2\pi/l_i$,
or in scaled variables 
$\tilde k_{li}=2\pi(4\mu/\hbar\omega_i)^{2/3}$,
for which (\ref{result}) is valid.
For the cigar shaped condensate of \cite{7}
this condition is in radial direction
($\tilde k_{lr} \approx 140$)
more limiting than in axial direction
($\tilde k_{la} \approx 770$).
Both limiting values place no restrictions on
the experimental observation 
for the low lying levels,
however for higher lying levels they may be relevant.
Let us consider the case of axial symmetry somewhat more generally:
For a level with quantum numbers $(n,j,m_l)$
in the notation of \cite{fliesser}
relation (\ref{result}) can be observed best 
in the range of the first maximum $k_{\mbox{\scriptsize max}}$ 
of the Bessel function $j_{n+m+1}(k)$, 
scaling as 
$k_{\mbox{\scriptsize max}} \sim (n+m_l)$ for large $n,m_l$.
If we have $n\gg m_l$, 
then $\omega\sim n$ 
and both limiting conditions 
on the energy $\omega_{njm} \ll \mu$ 
and on the momentum $k_{\mbox{\scriptsize max}} \ll k_l$ 
roughly coincide.
However if $n\ll m_l$, we have $\omega\sim m_l^{1/2}$
and the condition $k_{max}\gg k_l$  is the more limiting
one.

As an alternative experiment
to determine the collective excitations
and to observe (\ref{result})
one might also think of
inelastic scattering of light
\cite{griffin,fetter2}.
Unfortunately, this turns out to be hardly feasible,
as we now show.
Light scattered from an atomic gas
off resonance with any atomic transition
couples to the number density of atoms.
The cross section of elastic scattering of light
is proportional to the square of the spatial Fourier-transform 
of the equilibrium density $|\phi_0({\bf r})|^2$.
For an inhomogenous system elastic scattering occurs
in a finite angle, fixed by the condensate size.
Density fluctuations
can be seen as inelastic scattering of light.
The cross section of light shifted in frequency by $\omega$
via scattering from the density fluctuation $\phi$
is again proportional
to the Fourier-transform 
of this mode function
\cite{griffin}. One might hope to use this to test relation 
(\ref{result}).
For scattering transfering the wave vector ${\bf k}$
and at $T=0$ we get as a ratio of elastic to inelastic 
cross section
$(N\,\mu/\hbar\omega)
\times|\tilde\rho_0({\bf k})/\tilde\phi({\bf k})|^2$,
which unfortunately turns out as particularly unfavorable 
both in the limit $\hbar\omega \ll \mu$
and in the limit of large particle numbers.

The larger the transferred momentum ${\bf k}$, 
the better inelastic scattering can be distinguished
from the background of elastic scattering.
The maximum momentum, for which (\ref{result}) is valid,
is the momentum $\tilde k_l$ 
determined by the boundary width.
For present experiments \cite{4,7} this is  
by a factor $2-5$ smaller
than the largest momentum
transferable by light scattering,
occuring in the case of back-scattering.
Therefore the inelastic cross section from (\ref{result}) 
describes scattering for not too large angles of deflection only.
For the maximum momentum permissible by our approximations
the elastic cross section is of the order of $1/\tilde k_l^6$,
which follows from the (on this scale) discontinuous 
first derivative of $|\phi_0|^2$ at the boundary.
For small quantum numbers the inelastic cross section from
(\ref{result})
scales as $1/\tilde k_l^4$.
So collecting the prefactors we get as a minimum overall ratio
from elastic to inelastic scattering
$N\, (\mu/\hbar\omega)(\hbar\omega_i/4\mu)^{2/3}$
for inelastic scattering from level $\omega$
in direction $i$.
For the experiments of \cite{7} this ratio exceeds $10^4$
for all directions for the low lying levels.
This unfavorable ratio probably
precludes the possibility 
to test relation (\ref{result})
by inelastic light scattering.

In conclusion we have shown 
that the scaled Fourier-transform of the
hydrodynamic mode functions 
factorizes in a radial part and an angular part, even for harmonic
traps of arbitrary anisotropy.
The radial part is fixed by only one quantum number
and has the simple explicit form (\ref{result}).
This result was derived from a previously found conservation law.
The eigenvalues and eigenfunctions of the conserved operator
$\hat B$ determine completely the radial mode function in scaled
Fourier space .
We have discussed the range of validity of this property
and have suggested how it could be observed
in experiments using phase-contrast imaging.

M.F.\ thanks H.\ Schomerus for a valuable discussion.
This work has been supported by the Deutsche
Forschungsgemeinschaft through the Sonderforschungsbereich 237
''Unordnung und gro{\ss}e Fluktuationen''.

%

\end{document}